\documentclass[aps,graphicx,showpacs,twocolumn]{revtex4}%

\usepackage[dvipdfm]{graphicx}

\begin{document}

\title{Efficient generation of NOON states on two microwave-photon  resonators\footnote{Published in Chin. Sci. Bull.  \textbf{59}, 2829--2834 (2014)}}

\author{Ming Hua, Ming-Jie Tao, and
Fu-Guo Deng\footnote{{Corresponding author: fgdeng@bnu.edu.cn.}}}
\address{Department of Physics, Applied Optics Beijing Area Major Laboratory,
Beijing Normal University, Beijing 100875, China}


\date{\today }

\begin{abstract}
We present an efficient scheme for the generation of NOON states of
photons in circuit QED assisted by  a superconducting charge qutrit.
It is completed with two kinds of manipulations, that is, the
resonant operation on the qutrit and the resonator, and the
single-qubit operation on the qutrit, and they both are
high-fidelity operations. Compared with the one by a superconducting
transmon qutrit proposed by Su et al. (Sci. Rep. \textbf{4}, 3898
(2014)), our scheme does not require to maintain the qutrit in the
third excited state with a long time, which relaxes the difficulty
of its implementation  in experiment. Moreover, the level
anharmonicity of a charge qutrit is larger and it is better for us
to tune the different transitions of the charge qutrit resonant to
the
resonator, which makes our scheme faster than others.\\

\textbf{Key words:} entanglement production, NOON states,
microwave-photon resonators, superconducting charge qutrit, circuit
QED

\end{abstract}

%


\pacs{03.67.Bg,  42.50.Pq, 85.25.Cp}

\maketitle




\section{Introduction}

Quantum information is an important branch of quantum physics. It
includes mainly quantum communication and quantum computation
\cite{book,ren1,ren2,hong,long}. By far, many interesting quantum systems have been
presented for quantum information processing, such as nuclear
magnetic resonance \cite{longjcp,longprl}, quantum dots
\cite{QD1,QD3,QD4,QD6,QD9,QD12}, diamond nitrogen vacancy (NV)
centers \cite{NV1,NV2,NV3,NV4,NV5,NV6}, photonic systems
\cite{photon1,photon2,photon3,photon4,photon5}, circuit quantum
electrodynamics (QED)
\cite{Wallraff,liuyx,suncp,longglpra,longglnjp,a1,a2,a3,a4,a5,a6}, and so on.
Due to the good scalability \cite{Lucero} and convenient operation
on superconducting qubits, circuit QED has attracted much attention
in recent years.

Composed of the superconducting circuit and the superconducting 1D
resonator, circuit QED \cite{Blais} has some good characters for
completing  quantum information processing. The superconducting
circuit can act as a qubit perfectly.  The energy-level structure of
the qubit can be divided into $\Xi $, $\Lambda $, $V$, and $\Delta $
types \cite{You} which can not be found in atom systems. A relative
long life time of a superconducting qubit has been realized to reach
$0.1$ ms \cite{Rigetti}. The strong coupling strength between a
superconducting qubit and a superconducting resonator
\cite{Wallraff} has been demonstrated in the experiment.  All these
characters make circuit QED as a good platform for the quantum
computation based on superconducting qubits.  In 2009, DiCarlo
\emph{et al.} demonstrated a two-qubit algorithms with a
superconducting quantum processor \cite{DiCarlo}.  In 2012,  Reed
\emph{et al.} realized a three-qubit quantum error correction with
superconducting circuits \cite{Reed}, and in the same year,  Lucero
\emph{et al.} computed the prime factors with a Josephson phase
qubit quantum processor \cite{Lucero}  in which  they integrated
five superconducting resonators and four superconducting qubits in a
quantum processor.

A superconducting resonator can act as a cavity and a quantum bus,
which can be coupled to the distant qubits. The quality of the
resonator can be reached to $10^6$ and even $10^{12}$ \cite{Reagor},
that is, the superconducting resonators can also afford a powerful
platform for quantum information processing. In 2007, Schuster
\emph{et al.} resolved the photon number states in a superconducting
circuit \cite{Schuster}. In 2010,  Johnson \emph{et al.} realized a
quantum non-demolition detection of single microwave photons in a
circuit \cite{Johnson}, and in the same year,  Strauch \emph{et al.}
presented a method to synthesize an arbitrary quantum state of two
superconducting resonators \cite{Strauch}. In 2012, Strauch proposed
an all-resonant control of superconducting resonators with a drive
field \cite{FWStrauch}. In 2013,  we proposed a selective-resonance
scheme to perform a fast quantum entangling operation for quantum
logic gates on superconducting qubits \cite{huaSR}, assisted by one
or two superconducting resonators. By combination of the selective
resonance and the tunable period relation between a wanted quantum
Rabi oscillation and an unwanted one besides the positive influence
from the non-computational third levels of the superconducting
qubits, these universal quantum gates are significantly faster than
previous proposals and do not require any kind of drive fields.

Recently, the generation of the NOON state \cite{zhoulan}  on two resonators
attracted much more attention. In 2010, Strauch,  Jacobs, and
Simmonds \cite{Strauch} proposed a scheme  for completing the
generation of the NOON state on two resonators without using the
third non-computational excited energy level. The superconducting
qubit was operated with a selective rotation by using a drive field
whose amplitude should much smaller than the photon-number-dependent
Stark shifts on the qubit. That is, the operation time of the qubit
should be extended a little longer. In 2010, Merkel  and  Wilhelm
\cite{NJP2010} proposed a theoretic scheme for generating NOON
states on two resonators by using two superconducting qubits and
three superconducting resonators.  In 2011,  Wang \emph{et al.}
\cite{Wang} demonstrated  Merkel-Wilhelm scheme in experiment.   In
Ref.\cite{FWStrauch}, a novel method was proposed to generate the
NOON state on two resonators by using a complicated classical
microwave pulse and an all-resonant manipulation. It can get a very
high-fidelity NOON state within a much shorter time without using
the non-computational excited energy level. In 2013, Su \emph{et
al.} \cite{arxyang} proposed an interesting scheme for the
generation of the NOON state on two resonators with the resonant
operation between the transmon qubit and the superconducting
resonator, assisted by the single-qubit rotation. The scheme can be
completed with $2N$ steps, and in the first $N$ steps, the qubit
should be maintained in the third-excited state corresponding to the
case that the photon number in each resonator is zero.

In this paper, we proposed a scheme to produce the NOON state on two
resonators in a quantum processor composed of two tunable
superconducting resonators coupled to a tunable $\Xi$-type
three-energy-level superconducting qutrit.  Our scheme requires two
kind of quantum operations. One is the resonant operation on the
superconducting qutrit and the resonators. The other is the
single-qubit manipulation which can be completed by   applying  a
drive field on the qutrit. Our scheme can be used to produce the
NOON state on two resonators effectively in a simple and fast way,
compared with Merkel-Wilhelm scheme.  Moreover, it does not require
us to remain the qutrit in the third-excited state all the time,
which relaxes largely the requirements of its implementation in
experiment, compared with the previous work in Ref. \cite{arxyang}.

\section{Generation of the NOON state on two microwave-photon superconducting resonators}

\begin{figure}[h]
\par
\begin{center}
\includegraphics[width=8.2cm,angle=0]{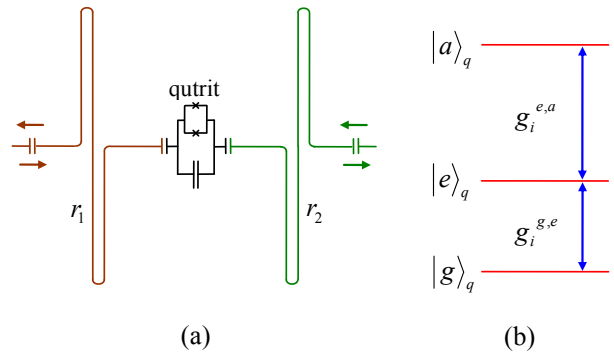}
\end{center}
\caption{(a) Schematic diagram  for generating  the NOON state on
two microwave-photon-resonator qudits. There are two resonators
coupled to a superconducting qutrit. The transition frequencies of
the qutrit and the resonators are tunable. (b) The structure for the
energy levels of a charge qutrit. $r_1$ and $r_2$ are the two
microwave-photon resonators. $g_i^{g,e}$ ($g_i^{e,a}$) is the
coupling strength between the resonator $r_i$ and the
superconducting qutrit in the transition between the states $\vert
g\rangle_q$ and $\vert e\rangle_q$ ($\vert e\rangle_q$ and $\vert
a\rangle_q$). } \label{fig1}
\end{figure}

Let us  consider a quantum system composed of two superconducting
resonators coupled to a superconducting qutrit,  shown in Fig.
\ref{fig1} (a).  The energy-level structure of the qutrit is the
$\Xi$ type,  which can be found in a superconducting charge qubit,
shown in Fig. \ref{fig1} (b). In order to construct the NOON state
on the two resonators $r_1$ and $r_2$, we exploit the lowest three
energy levels of the qutrit, denoted by $\left\vert
g\right\rangle_q$, $\left\vert e\right\rangle_q$, and $\left\vert
a\right\rangle_q$ with the energy $E_{g}<E_{e}<E_{a}$. The
Hamiltonian of the system composed of the two resonators and the
qutrit is (under the rotating-wave approximation, and we choose
$\hbar =1$ below)
\begin{eqnarray}       
H &=&\sum\limits_{l=g,e,a}E_{l}\left\vert l\right\rangle
_{q}\left\langle l\right\vert   + \sum_{i=1,2}[\omega
^{r_i}a_{i}^{+}a_{i}+g_{i}^{g,e}(a_{i}^{+}\sigma
_{g,e}^{-}\nonumber\\&& +a_{i}\sigma _{g,e}^{+}
+g_{i}^{e,a}(a_{i}^{+}\sigma _{e,a}^{-}+a_{i}\sigma _{e,a}^{+})].
\label{H}
\end{eqnarray}
Here, $\omega^{r_i}$ and $a_{i}^{+}$ are the transition frequency
and the creation operator of the resonator $r_i$, respectively.
$\sigma_{g,e}^{+}$ and $\sigma_{e,a}^{+}$ are the creation operators
of the two transitions $\left\vert
g\right\rangle_{q}\rightarrow\left\vert e\right\rangle_{q}$ and
$\left\vert e\right\rangle_{q}\rightarrow\left\vert
a\right\rangle_{q}$ of the qutrit, respectively. $g_{i}^{g,e}$ is
the coupling strength between the resonator $r_i$ and the qutrit in
the two transitions  $\vert g\rangle_q$ and $\vert e \rangle_q$, and
$g_{i}^{e,a}$ is the coupling strengths between the resonator $r_i$
and the qutrit in the two transitions  $\vert e\rangle_q$ and $\vert a
\rangle_q$.

In order to turn on or off the interaction between the resonators
and the qutrit, on one hand, one can tune the transition frequency
of the qutrit by using the external magnetic flux, or tune the
transition frequency of the resonator to make them resonate or
largely detune with each other. On the other hand, one can tune  the
coupling strength between the qutrit and the resonator. It worth
noticing that a tunable resonator \cite{Sandberg} and a tunable
coupling qubit \cite{Allman,Bialczak} have been demonstrated in
experiment.

The principle of our scheme for generating the NOON state on two
microwave-photon resonators efficiently is shown in Fig.
\ref{fig1}(a). Suppose the initial state of the system is
\begin{eqnarray}         
\left\vert \phi \right\rangle  &=&\frac{1}{\sqrt{2}}(\left\vert
g\right\rangle_{q}+\left\vert e\right\rangle_{q})\otimes \left\vert
0\right\rangle _{1}\left\vert 0\right\rangle _{2}\nonumber\\
&=&\frac{1}{\sqrt{2}}(\left\vert g\right\rangle_{q} \left\vert
0\right\rangle _{1}\left\vert 0\right\rangle _{2}+\left\vert
e\right\rangle_{q} \left\vert 0\right\rangle _{1}\left\vert
0\right\rangle _{2}). \label{initial}
\end{eqnarray}
Here the subscripts $1$ and $2$ represent the two resonators $r_1$
and $r_2$, respectively.  That is, the qutrit is in the state
$\frac{1}{\sqrt{2}}(\left\vert g\right\rangle_{q} +\left\vert
e\right\rangle_{q} )$, and the resonators are in the state
$\left\vert 0\right\rangle_{1}\left\vert 0\right\rangle _{2}$. Here
and below, $\vert n\rangle_i$ is the Fock state of the resonator
$r_i$, which means there are $n$ microwave photons in the resonator
$r_i$ ($i=1,2$).  To generate the NOON state \cite{Strauch}
\begin{eqnarray}            
\left\vert\phi\right\rangle_{NOON} =\frac{1}{\sqrt{2}}(\left\vert
N\right\rangle _{1}\left\vert 0\right\rangle _{2}+\left\vert
0\right\rangle _{1}\left\vert M\right\rangle _{2}) \label{noon}
\end{eqnarray}
on  $r_1$ and $r_2$ ($N=M$ is a special situation of the NOON state), our scheme needs $N+M$ steps. The first $N$
steps are described as follows.

Step $1$: By making both $r_1$ and  $r_2$  detune largely  with the
qutrit, one can use a drive field with the frequency equivalent to
the transition frequency $\omega_{e,a}$ of the qutrit to pump the
state of the qutrit  from  $\left\vert e\right\rangle_{q}$ to
$\left\vert a\right\rangle_{q}$.  The amplitude of the drive field
is chosen with a proper value for avoiding to pump the state from
$\left\vert g\right\rangle_{q}$ to $\left\vert e\right\rangle_{q}$.
Here $\omega_{e,a}\equiv E_{a}-E_{e}$. After the operation time
$\Omega _{e,a}t=\pi$ ($\Omega_{e,a}$ is the proper amplitude of the
drive field for pumping the qutrit from $\left\vert
e\right\rangle_{q}$ to $\left\vert a\right\rangle_{q}$), the state
of the system evolves into
\begin{eqnarray}        
\frac{1}{\sqrt{2}}(\left\vert g\right\rangle _{q}\left\vert 0\right\rangle
_{1}\left\vert 0\right\rangle _{2}-i\left\vert a\right\rangle _{q}\left\vert
0\right\rangle _{1}\left\vert 0\right\rangle _{2}).
\label{1}
\end{eqnarray}
Subsequently, one can  tune the transition frequencies of the qutrit
and the two resonators to make  $r_1$ resonate with  the qutrit in
the transition $\left\vert
e\right\rangle_{q}\leftrightarrow\left\vert a\right\rangle_{q}$. If
the coupling strength between
 $r_1$ and the qutrit is tuned with a proper value before the
resonance, one can neglect the interaction between  $r_1$ and  the
qutrit in the transition $\left\vert
g\right\rangle_{q}\leftrightarrow\left\vert e\right\rangle_{q}$.
Meanwhile,  $r_2$ and the qutrit detune largely  with each other.
After the interaction time $g_{1}^{e,a}t=\pi$, the state of the
system becomes
\begin{eqnarray}         
\frac{1}{\sqrt{2}}(\left\vert g\right\rangle _{q}\left\vert 0\right\rangle
_{1}\left\vert 0\right\rangle _{2}-\left\vert e\right\rangle _{q}\left\vert
1\right\rangle _{1}\left\vert 0\right\rangle _{2}).
\label{11}
\end{eqnarray}

Step $j$ ($j=2,3,...,N$): By repeating the operation of the step 1
for  $N-1$ times and maintaining  $r_2$  detuning largely with $r_1$
and the qutrit all the time,  the state of the system is changed to
be
\begin{eqnarray}         
\frac{1}{\sqrt{2}}(\left\vert g\right\rangle _{q}\left\vert 0\right\rangle
_{1}\left\vert 0\right\rangle _{2}+(-1)^{N}\left\vert e\right\rangle _{q}\left\vert
N\right\rangle _{1}\left\vert 0\right\rangle _{2}).
\label{N1}
\end{eqnarray}
The whole operation time is
\begin{eqnarray}        
t = t_{d}+t_{r}. \label{N}
\end{eqnarray}
Here, $t_{d}=\sum_N \frac{ N\pi }{\Omega _{e,a}}$ is the
rotated-operation time of the qutrit and  $t_{r}=\sum_N\frac{\pi
}{2g_{i}^{e,a}\sqrt{N}}$ is the resonated-operation time between the
qutrit and the $r_1$.

The details of the first $N$ steps have been described above. The next $M$ steps are described as follows.

Step $1'$: By making both $r_1$ and $r_2$, detune largely with the
qutrit, one can apply a drive field with the frequency equivalent to
the transition frequency $\omega_{g,e}$ of the qutrit to rotate the
states of the qutrit with $\left\vert
g\right\rangle_{q}\leftrightarrow\left\vert e\right\rangle_{q}$. By
choosing the proper amplitude of the drive field, one can avoid to
flip the qutrit with $\left\vert
e\right\rangle_{q}\leftrightarrow\left\vert a\right\rangle_{q}$.
After the operation time $\Omega _{g,e}t=\pi $, the state of the
system evolves from Eq.(\ref{N1}) to
\begin{eqnarray}         
\frac{1}{\sqrt{2}}(-i\left\vert e\right\rangle _{q}\left\vert 0\right\rangle
_{1}\left\vert 0\right\rangle _{2}-(-1)^{N}i\left\vert g\right\rangle _{q}\left\vert
N\right\rangle _{1}\left\vert 0\right\rangle _{2}).
\label{M+1}
\end{eqnarray}
Applying a drive field with the frequency equivalent to
the transition frequency $\omega_{e,a}$ of the qutrit, one can pump the
state of the qutrit from $\left\vert e\right\rangle_q$ to $\left\vert
a\right\rangle_q$. The amplitude of the drive field is chosen with a
proper value for avoiding to pump the state from $\left\vert
g\right\rangle_q$ to $\left\vert e\right\rangle_q$. After the operation
time   $\Omega _{e,a}t=\pi $, the state of the system evolves into
\begin{eqnarray}         
\frac{1}{\sqrt{2}}(-\left\vert a\right\rangle _{q}\left\vert 0\right\rangle
_{1}\left\vert 0\right\rangle _{2}-(-1)^{N}i\left\vert g\right\rangle _{q}\left\vert
N\right\rangle _{1}\left\vert 0\right\rangle _{2}).
\label{M+11}
\end{eqnarray}
Subsequently, one can tune the transition frequencies of the qutrit
and the two resonators to make  $r_2$ resonate with  the qutrit in
the transition $\left\vert e\right\rangle_{q}\rightarrow\left\vert
a\right\rangle_{q}$. If the coupling strength between t  $r_1$ and
the qutrit is tuned with a proper value before the resonance, one
can neglect the interaction between  $r_2$ and  the qutrit in  the
transition $\left\vert g\right\rangle_{q}\rightarrow\left\vert
e\right\rangle_{q}$. Meanwhile,  $r_1$ and the qutrit
 detune largely with each other. After the interaction time
$g_{2}^{e,a}t=\pi$, the state of the system becomes
\begin{eqnarray}         
\frac{1}{\sqrt{2}}(i\left\vert e\right\rangle _{q}\left\vert 0\right\rangle
_{1}\left\vert 1\right\rangle _{2}-(-1)^{N}i\left\vert g\right\rangle _{q}\left\vert
N\right\rangle _{1}\left\vert 0\right\rangle _{2}).
\label{M+12}
\end{eqnarray}

Step $j'$ $(j'=2,3,...,M-1)$: By repeating the operation of the step
$1'$ for $M-2$ times, and maintaining  $r_1$ detuning largely with
the qutrit all the time,  the state of the system is changed to be
\begin{eqnarray}         
\frac{1}{\sqrt{2}}((-1)^{M-1}i\left\vert e\right\rangle
_{q}\left\vert 0\right\rangle _{1}\left\vert M-1\right\rangle _{2}
-(-1)^{N}i\left\vert g\right\rangle _{q}\left\vert N\right\rangle
_{1}\left\vert 0\right\rangle _{2}). \nonumber\\ \label{M+13}
\end{eqnarray}

The final step: Applying a single-qubit operation to complete the
rotations of the states  $(-1)^{M-1}i\left\vert
e\right\rangle_{q}\rightarrow i\left\vert e\right\rangle_{q}$ and
$(-1)^{N-1}i\left\vert g\right\rangle _{q}) \rightarrow\left\vert
g\right\rangle _{q}$, the state of the system evolves into
\begin{eqnarray}         
\frac{1}{\sqrt{2}}(i\left\vert e\right\rangle
_{q}\left\vert 0\right\rangle _{1}\left\vert M-1\right\rangle
_{2}+\left\vert g\right\rangle _{q}\left\vert
N\right\rangle _{1}\left\vert 0\right\rangle _{2}). \label{M+14}
\end{eqnarray}
By resonating  $r_2$ and the qutrit in  the transition $\left\vert
g\right\rangle_{q}\leftrightarrow\left\vert e\right\rangle_{q}$, and
making   $r_1$ detune largely with the qutrit, the state shown in
Eq.(\ref{M+14}) is changed to be
\begin{eqnarray}            
\frac{1}{\sqrt{2}}(\left\vert N\right\rangle
_{1}\left\vert 0\right\rangle _{2}+\left\vert 0\right\rangle _{1}\left\vert
M\right\rangle _{2})\otimes\left\vert g\right\rangle _{q}.
\label{M}
\end{eqnarray}
Here, we have generated the NOON state on two
microwave-photon resonators efficiently. The operation time of the
second $M$ steps is
\begin{eqnarray}        
t' = t'_{d}+t'_{r}.
\label{M-1}
\end{eqnarray}
$t'_{d}=\sum_M\frac{M\pi }{\Omega _{e,a}}$ is the
rotation-operation time of the qutrit and
$t'_{r}=\sum_{M}\frac{\pi }{2g_{i}^{e,a}\sqrt{M}}$ is the
resonance-operation time between the qutrit and  $r_2$. In which,
we neglect the operation time of the single-qubit operation in
the final step for generating the NOON state with large number of the
$N$ and $M$.

\section{DISCUSSION AND SUMMARY}

We have described the process of our scheme for generating the NOON
state on two superconducting resonators which are coupled to a
$\Xi$-type-energy-level structure superconducting charge qutrit. It
includes two kinds of quantum operations. The first one is the
resonant operation on the qutrit and the resonators. The second one
is  the single-qubit operation on the qutrit. They are the
high-fidelity, high-efficiency, and simple quantum operations in
experiment in circuit QED systems. The whole operation time of our
scheme for generating the NOON state
$\left\vert\phi\right\rangle_{noon}$ is
\begin{eqnarray}         
T &=& \sum_{j=1}^{N} \left(\frac{ j\pi }{\Omega _{e,a}} + \frac{\pi
}{2g_{i}^{e,a}\sqrt{j}}\right) + \sum_{j'=1}^{M}
\left(\frac{j'\pi}{\Omega _{e,a}}+\frac{\pi
}{2g_{i}^{e,a}\sqrt{j'}}\right).\nonumber\\ \label{time}
\end{eqnarray}
In the calculation for the operation time in our scheme, we neglect
the time for changing the transition frequencies of the
superconducting qutrit and the superconducting resonator, and the
operation time of the single-qubit operation in the final step.

Compared with  the one in Ref.\cite{Strauch}, our scheme  for
generating the NOON state on superconducting resonators is  much
faster  as it is composed of the resonant controls. Compared with
the one in Refs.\cite{NJP2010,Wang}, both the number of the
resonators and that of the qutrits required in our scheme are much
smaller as there are three superconducting resonators and two
superconducting qutrits in the scheme in Refs.\cite{NJP2010,Wang},
but only two superconducting resonators and a superconducting qutrit
used in our scheme. Moreover, the single-qubit operation required in
our scheme can be achieved with the simple classical drive field,
and it is simpler than the one used in Ref.\cite{FWStrauch} as the
amplitude of the drive field should be designed with a complex type
and it is difficult to be realized in experiment in the latter. In
Ref. \cite{arxyang}, a similar method is used to generate the NOON
state on two resonators. In their work, the transmon qutrit should
be maintained in the first $N$ steps in the third excited state when
there is no microwave photons in each resonators. It worth noticing
that the higher excited states lead  to a lower fidelity operation
\cite{FWStrauch}. Luckily, our scheme does not require us to
maintain the qutrit in its third excited state all the time, which
relaxes the requirements of its implementation in experiment,
compared with  the one in Ref.\cite{arxyang}. Compared with a
transmon qutrit, the level anharmonicity of a charge qutrit is
larger and it is better for us to tune the different transitions of
the charge qutrit resonant to the resonator \cite{chargegood}.

%
%

In summary, we have proposed an efficient scheme to generate the
NOON states on two superconducting resonators, assisted by a
superconducting qutrit.  It requires some high-fidelity quantum
operations, that is, the resonant operation on the qutrit and the
resonator and the single-qubit operation on the qutrit. Our scheme
is a fast and simple one. Moreover, it does not require to maintain
the qutrit in the third excited state with a long time, which
relaxes the requirements of its implementation in experiment.

\section*{Acknowledgements}

This work is supported by the National Natural Science Foundation of
China under Grant No. 11174039 and NECT-11-0031.


\end{document}